\documentclass[article]{aa}
\usepackage{epsf}
\usepackage{times}
\begin{document}
\topmargin0.0cm
\thesaurus{06     % Form., struct., & evolut. of stars
 (08.02.3;     %  Stars: binaries: general
  08.05.3;     %  Stars: evolution
  08.09.3;     %  Stars: interiors
  08.16.7:PSR~J1012+5307;%  Pulsars: PSR~J1012+5307
  08.23.1)     %  Stars: white dwarfs 
}
\title{The evolution of helium white dwarfs:\\
I. The companion of the millisecond pulsar PSR~J1012+5307\thanks{Tables 5-11 are
only available in electronic form at the CDS via ftp 130.79.128.5}}
%\subtitle{}
\author{T.\ Driebe\inst{1,2}, D.\ Sch\"onberner\inst{2
},
T.\ Bl\"ocker\inst{1,3}, F.\ Herwig\inst{2}}%
\offprints{T. Driebe}
\institute{
  Max-Planck-Institut f\"ur Radioastronomie, Auf dem H\"ugel 69, D-53321 Bonn,
  Germany \\ (driebe@speckle.mpifr-bonn.mpg.de; bloecker@speckle.mpifr-bonn.mpg.de)
\and
  Astrophysikalisches Institut Potsdam, An der Sternwarte 16, D-14482 Potsdam,
  Germany \\ 
  (deschoenberner@aip.de; fherwig@aip.de)
\and
  Institut f\"ur Theoretische Physik und Astrophysik, Universit\"at Kiel,
  D-24098 Kiel, Germany % (bloecker@astrophysik.uni-kiel.de)
}
\date{Received date /  accepted date}
\maketitle
\markboth{T. Driebe et al.:  Evolution of He-white dwarfs}{} 
%%%%%
%%%%%
\begin{abstract}
%%%%% 
%%%%% 
%%%%% 

We present a grid of evolutionary tracks for low-mass white dwarfs
with helium cores in the mass range from $ 0.179$ to $0.414~{\rm M}_{\odot}$.
The lower mass limit is well-suited for comparison with white dwarf companions
of millisecond pulsars. The tracks are
based on a $1~{\rm M}_{\odot}$ model sequence extending from the pre-main 
sequence stage   %%%%the Hayashi limit 
up to the tip of the red-giant branch.
Applying large mass loss rates   
at appropriate positions forced
the models to move off the giant branch. The further
evolution was then followed across the Hertzsprung-Russell diagram 
and down the cooling branch. At maximum effective temperature
the envelope masses above the helium cores increase from 0.6 to
$5.4\cdot 10^{-3}~{\rm M}_{\odot}$ for decreasing mass.
We carefully checked for the occurrence of
thermal instabilities of the hydrogen shell by adjusting the
computational time steps accordingly. Hydrogen flashes have been
found to take place only in the mass interval $0.21\la M/{\rm M}_{\odot}\la 0.3$.

The models show that hydrogen shell burning contributes significantly to 
the luminosity budget of white dwarfs with helium cores.
For very low masses the hydrogen shell luminosity remains to be  
dominant even down to effective temperatures 
well below $10\,000\,{\rm K}$. Accordingly, the
corresponding cooling ages are significantly larger
than those gained from model calculations which neglect
nuclear burning or the white dwarf progenitor evolution.

Using the atmospheric parameters of the white dwarf in the
\object{PSR~J1012+5307} system we determined a mass
of $M=0.19\pm 0.02~{\rm M}_{\odot}$ and a cooling age
of $6\pm 1 \mathrm{Gyr}$, in good agreement with
the spin-down age, 7 Gyr, of the pulsar.
 \keywords{Stars: evolution --  
           Stars: interiors --
           White dwarfs -- 
           Binaries: general --
           Pulsars: PSR~J1012+5307 
         }
%_____________________________________ Do not leave a blank line here!
\end{abstract}

\section{Introduction} \label{intro}
The theory of stellar evolution 
shows that white dwarfs (WD) with a carbon-oxygen core (CO-WDs) can be well
understood in terms of the evolution of single stars  
with initial masses between 1 and about $6\dots 8\,{\rm M}_{\odot}$.
After central hydrogen burning,  
stars in this mass range develop first a helium core surrounded by 
a hydrogen-burning shell (red-giant branch or RGB) and later, after 
completion of central helium burning, a compact carbon-oxygen core
 surrounded by two %%%nuclear-active
shells burning hydrogen and helium, respectively (asymptotic giant branch
or AGB). At high luminosities strong stellar winds erode
the stellar envelope effectively. When
the envelope mass drops below a few $10^{-2}~{\rm M}_{\odot}$ the model
leaves the AGB, becomes a central star of a planetary nebula
and finally cools down as a white dwarf.
The final masses 
range typically between about 0.5 and $1~{\rm M}_{\odot}$
(see e.g.\ Iben \& Renzini \cite{IR83}; 
Sch\"onberner \cite{Sch79}, \cite{Sch83}; Vassiliadis \& Wood \cite{VW93}, 
\cite{VW94}; Bl\"ocker \cite{Ba}, \cite{Bb}).

White dwarfs with smaller masses cannot be produced by single-star
evolution since their progenitors would have initial masses below 
$0.5~{\rm M}_{\odot}$ and thus do not finish their main sequence evolution 
within a Hubble time.
Instead one has to invoke mass transfer in a close binary system 
whereby
the donor's evolution towards central helium ignition is choked off 
-- the so-called case B mass exchange (Kippenhahn \& Weigert \cite{KW}). The
remnant is a low-mass  object ($ M < 0.5~{\rm M}_{\odot}$), consisting of a helium
core with a hydrogen-rich envelope still burning hydrogen at its bottom, 
which contracts slowly
towards a white dwarf configuration, quite similar to the more massive 
and more luminous remnants from the AGB. 

Two known common types of binary systems can
contain such helium white dwarfs (He-WDs). 
Firstly the so-called double-degenerate
systems, where both components are white dwarfs (see e.g. Saffer et al. \cite{SLY}).
Examples for such systems with one or even two He-WDs
are \object{WD\,0957-666} (Bragaglia et al. \cite{BGRO}; Moran et al. \cite{MMB}),
 \object{WD\,0135-052} (= \object{L870-2} = \object{EG11}, 
Saffer et al. \cite{SLO}), \object{PG\,1101+364} (= \object{Ton 1323},
Marsh 1995)
and the five systems studied by Marsh
et al. (\cite{MDD}).
The second type are the so-called millisecond pulsar
systems (hereafter MSP), where the He-WD is the
companion of a pulsar with a high rotational speed in a
nearly circular orbit. 

So far about three dozens of these systems are known.
Most of them have a low-mass white dwarf companion (see for example
Bailes \& Lorimer 1995, Camilo et al. 1996, Ray et al. 1996, Lyne \cite{Ly1}).
Reviews of MSP properties are given, e.g., by Verbunt (\cite{V}), 
Shore et al. (1994), Phinney 
\& Kulkarni (\cite{PK}), Rappaport et al. (\cite{RPJSH}), 
Nicastro et al. (\cite{NLLHBS}), Camilo (\cite{C}) or Lyne (\cite{Ly}).
The evolution of the progenitor
system can be explained by highly non-conservative mass
exchange events (common envelope evolution).
A detailed description of the binary scenario which leads to the formation 
of a MSP system has recently been given, e.g., by Tauris \& Bailes (\cite{TB}), 
Tauris (\cite{T}) or Ergma \& Sarna (\cite{ES}). In progenitor systems to MSPs 
the more massive component ($M_1\approx 10~{\rm M}_{\odot}$) fills its
Roche volume, looses a substantial amount of mass but follows the evolution
towards a supernova explosion with the formation of a neutron star
which might be detected as a radio pulsar. 
The less massive secondary component ($M_2\approx 1~{\rm M}_{\odot}$) continues 
its evolution, evolves off the main sequence to become a red giant and will 
eventually also exceed its Roche lobe to spill its mass onto 
the neutron star. This accretion of matter spins up the 
neutron star to rotational periods of the order of milliseconds.
In this phase the system can be identified as a low mass X-ray 
binary system (LMXB) like \object{Cyg X-2} or \object{2S 0921-63}
(Webbink et al. \cite{WRS}; Verbunt \cite{V};
for a population synthesis of LMXBs see Iben et al. \cite{ITY95}).
Continuing mass exchange of the secondary causes its envelope  mass to fall 
below a critical value.  The system becomes detached again, and the 
secondary evolves to a white dwarf with a helium core while the
pulsar's rotation slows down again by energy loss at the expense of rotational
energy and the pulsar properties are activated again.

Using smaller initial mass ratios the scenarios for the emergence of
MSs also describe, in principle,
the formation of double degenerate systems, i.e. systems containing
two white dwarfs (CO+CO, He+CO, He+He). 
For a discussion of these systems see for example Webbink (\cite{W84}), 
Iben \& Tutukov (\cite{IT84a}, \cite{IT86}), Cameron \& Iben (\cite{CI}) 
and, more  recently, Sarna et al. (\cite{SMC}).

MSP systems open the unique
possibility to check the spin-down
theory of pulsars by independent age
determinations of the white dwarf
components once the atmospheric parameters of the latter are known.
\object{PSR~J1012+5307} is so far the only system where the white dwarf has been studied 
with sufficient precision as to allow for comparing spin-down and cooling ages 
(see e.g.\ Lorimer et al. \cite{LFLN}) and to determine the components' masses.
 
\object{PSR~J1012+5307} was identified
by Nicastro et al. (\cite{NLLHBS}). They found a 
rotational period $P_{\rm rot}=5.26~{\rm ms}$ and a compact companion with
a minimum mass of 0.11$~{\rm M}_{\odot}$ in an orbit with period  
$P_{\rm orb}=14.5\,{\rm h}$.
The surface parameters of the companion have been determined
by van Kerkwijk et al. (\cite{KBK}) and Callanan et al. (\cite{CGK}).
Due to different atmospheric models the derived parameters differ
somewhat: van Kerkwijk et al. (\cite{KBK}) found
$\log g=6.75\pm 0.07$ and $T_{\rm eff}=8550\pm 25\,{\rm K}$ while
Callanan et al. (\cite{CGK}) give
$\log g = 6.34\pm 0.20$ and $T_{\rm eff} = 8670\pm 300~{\rm K}$.
According to these atmospheric parameters the 
companion is a low-mass helium white dwarf as suggested by the theoretical 
scenarios sketched above.
Van Kerkwijk et al. (\cite{KBK}) estimated a mass of $0.16\pm 0.02 \,{\rm M}_{\odot}$
by extrapolating from more massive carbon-oxygen models. The same result
was found by Callanan et al. (\cite{CGK}) from their analysis and
white dwarf modelling.

Lorimer et al. (\cite{LFLN}) determined the 
spin-down age $\tau=7.0$~Gyr assuming $P_0\ll P_{\rm rot}$,
where $P_0$ is the initial pulsar period after the end of the spin-up phase.
The age of the  
MSP can be estimated as well by the cooling age of
the WD companion. However, the estimates based on
existing white dwarf calculations are controversial: while
Lorimer et al. (\cite{LFLN}) derived a cooling age for
the white dwarf of about 0.3~Gyr, which is
a factor 20 lower than the pulsar's spin-down age, 
Alberts et al. (\cite{ASH}) 
give a value of 7~Gyr, in full agreement with the spin-down
age of the pulsar.

To solve the problem of age determination of MSPs
two different strategies are possible: The first concerns the 
critical examination of the spin-up and spin-down theories for
pulsars. On one hand, an uncertainty in the spin-down age determination 
arises from the assumption $P_0\ll P$. Violations of this assumption
result in lower spin-down ages (see for example 
Camilo et al. 1994).
On the other hand, Burderi et al. (\cite{BKW}) showed that by
assuming an accretion-induced field decay 
instead of spontaneous field decay 
substantially lower spin-down ages can be obtained. 
Following their conclusions MSP systems would thus
have characteristic ages of a few $10^8$~years only.
In the particular case of \object{PSR~J1012+5307} Burderi et al. (\cite{BKW})
found a spin-down age of about $3.7 \cdot 10^8$~yr. 

The other strategy for deriving corresponding ages 
involves the reexamination of the
evolutionary models of low-mass white dwarfs.
Studies of such objects are rare,
in particular for $M<0.2~{\rm M}_{\odot}$ as is appropriate for 
\object{PSR J1012+5307}.
Several authors, i.e.\ Kippenhahn et al. (\cite{KKW}), Kippenhahn et al.
(\cite{KTW}), Refsdal \& Weigert (\cite{RW}), Gianonne et al. (\cite{GRW}),
Iben \& Tutukov (\cite{IT86}) and Castellani et al. (\cite{CLR})
computed models of He-WDs with $M>0.2~{\rm M}_{\odot}$ by simulating mass 
exchange in close binaries. Typical is that
all of these studies have encountered thermal instabilities 
(or hydrogen flashes) in models with 
$M > 0.25~{\rm M}_{\odot}$ due to unstable hydrogen burning on the cooling branch. 

Chin \& Stothers (\cite{ChSt}) and Webbink (\cite{W75}) constructed
He-WD models by following the evolution of single stars of the appropriate
low masses.\ Webbink (\cite{W75}) calculated a grid of
sequences in the mass range $ 0.1 < M/\,{\rm M}_{\odot}<0.5$.
However, due to large time steps used his tracks do not
show any thermal instabilities. 
From Webbinks calculations one can infer that the white dwarf companion in
the \object{PSR~J1012+5307} system should have an age of $\tau \approx 10$~Gyr.
Chin \& Stothers (\cite{ChSt}) did not include hydrogen shell burning in 
their white dwarf models. 
According to their $M=0.2~ {\rm M}_{\odot}$ track,
the white dwarf in the \object{PSR J1012+5307} system is only 0.3~Gyr old.
Hereafter, models with the explicit assumption 
$L_{\rm nuc}=0$ will be referred to as contraction models.

Alberts et al. (\cite{ASH}) calculated He-WD
models between  $0.17<M/\,{\rm M}_{\odot}<0.25$ by simulating
binary evolution with low-mass components. 
They found no thermal instabilities of the hydrogen burning shell.
Their models give for  the \object{PSR J1012+5307} companion's mass
and age $M=0.185~{\rm M}_{\odot}$ and $\tau=7$ Gyr, respectively.

Althaus \& Benvenuto (\cite{AB}) and Benvenuto \& Althaus (\cite{BA})
presented models in the mass range $0.15 < M/\,{\rm M}_{\odot}<0.5$.
Althaus \& Benvenuto (\cite{AB}) treated He-WDs without any hydrogen envelopes,
and thus did not allow for nuclear burning.
Benvenuto \& Althaus (\cite{BA}) and
Hansen \& Phinney (\cite{HPa}) ($0.1 < M/\,{\rm M}_{\odot}<0.5$)
started their calculations considering hydrogen burning,
but found it to be insignificant.
All sets of tracks of these authors 
display very similar cooling properties, and one can 
estimate a cooling age
of about $\tau \la 0.5$ Gyr for the white 
dwarf in the \object{PSR J1012+5307} system.
Note, that, however, their initial
models differ from those based on
calculations of RGB progenitors with mass loss. 

It is obvious from this comparison that the cooling properties
of low-mass white dwarfs are extremely sensitive to the initial
conditions. We suggest that initial models closest to 
real He-WD progenitors result from explicit  modeling
of their evolutionary history which determines
the model's envelope mass and thermomechanical structure.
Therefore we have calculated an extensive grid of model sequences
with full consideration of nuclear burning.
Our investigations revealed that the simultaneous consideration
of these two aspects, i.e. nuclear burning and the evolutionary
history, results in considerably longer evolutionary ages
of the white dwarfs. Consequences for the mass determination are
evident as well.
A full presentation of this evolutionary grid, with 
an extensive discussion of thermal instabilities and the
mass-radius relation is deferred to a forthcoming paper.
Here we will discuss only those model properties that are important
to interpret millisecond pulsar systems.

This  paper is organized as follows: in Sect.~\ref{Shewd} 
we describe the stellar evolution code and the method
used to calculate evolutionary models presented
in this paper. Results are given in Sect.~\ref{Sres} and applied
to the particular system \object{PSR J1012+5307}. 
Sect.~\ref{Sconcl} summarizes our results.   

\section{The evolutionary calculations}  \label{Shewd}
The evolutionary code used is essentially the one 
described by Bl\"ocker (\cite {Ba}),  with several modifications.
Nuclear burning is accounted for 
via a nucleosynthesis network including 30 isotopes with all important 
reactions up to carbon burning similar as in 
El Eid (\cite {El}). The most recent radiative opacities 
by Iglesias et al. (\cite{IRW}) and Iglesias \& Rogers (\cite{IR}) [OPAL], 
supplemented by those of Alexander \& Ferguson (\cite{AF}) in 
the low-temperature region, are employed. Diffusion is not considered.
The initial composition is 
$(Y,Z)=(0.28,0.02)$, the mixing length parameter
$\alpha=1.7$ follows from calibrating a solar model. 
The Coulomb corrections to the equation of state are those given by
Slattery et al. (\cite{SDW}). We note in passing that for the results presented
here the quality of the Coulomb correction treatment is of no significance.

For comparison we have also computed evolutionary sequences based on 
  different input physics:
\begin{itemize}   
  \item   sequences  using the older opacities of
          Cox \& Stewart (\cite{CSa}, \cite{CSb}, \cite{CS70}) [CS];   
          and 
  \item  sequences  using equilibrium nuclear reaction 
  rates for hydrogen and helium burning. 
 \end{itemize}
In the following we will only refer to the sequences based on the OPAL opacities
with full employment of the nuclear network if not mentioned differently.
\subsection{Method of calculation} \label{moc}
Because we primarily focused our study on the cooling
behaviour of low-mass white dwarfs and the implications for the  mass-radius 
relation we did not calculate the mass exchange phases
during the RGB evolution in detail (see for this subject,
e.g., Tauris \cite{T}, Tauris \& Bailes \cite{TB}).
Rather, for our purpose it was sufficient to simulate the mass-exchange 
episode by subjecting a RGB model to a sufficiently large mass loss rate 
(Iben \& Tutukov \cite{IT86}, Castellani et al. \cite{CLR}). We emphasize that,
once mass loss has been turned off, the model's further evolution does
not depend on the details of the previous mass loss episode.
The evolutionary speed across the Hertzsprung-Russell diagram
is rather controlled by the model's structural changes caused
by the actual envelope-mass reduction due to hydrogen burning and mass loss.

In order to get realistic starting models, we calculated  a
1~${\rm M}_{\odot}$ sequence from the pre-main sequence phase up to the tip of the RGB. 
Along the RGB we applied mass-loss rates $\dot{M}_{\rm R}$ according to 
Reimers (\cite{R}) with $\eta = 0.5$ as in Maeder \& Meynet (\cite{MAME}).
At appropriate positions high mass loss rates, $\dot{M}_{\rm high}$, 
were invoked in order to get models of desired final mass, $M$.
Since in the actual situations
Roche-lobe overflow is assumed to occur on a nuclear
time scale, the maximum applied mass loss rate was chosen in such a way as
not to destroy the model's thermal equilibrium.
Accordingly, $\dot{M}_{\rm high}$\ varied from $\dot{M}_{\rm high}$\ $\approx
10^{-9}~{\rm M}_{\odot}\,{\rm yr}^{-1}$ for  $M \approx 0.15~{\rm M}_{\odot}$ to about
$\dot{M}_{\rm high}$\ $\approx 10^{-6}\,{\rm M}_{\odot}\,{\rm yr}^{-1}$ for 
$M \approx 0.4~{\rm M}_{\odot}$. 
Below a critical envelope mass, $M_{\rm env}^{\rm crit}$, 
sufficient densities and temperatures
to continue hydrogen burning can only be maintained if the envelope
contracts. Then, the model starts to leave the RGB.
At this point of evolution mass loss was virtually switched off
by decreasing $\dot{M}$ over a short transition period until
$\dot{M}$\ = $\dot{M}_{\rm R}$\ was reached.
In general we chose
$T_{\rm eff}=5000$\ K to be the point where $\dot{M}$\ should have
reached the Reimers rate with $\eta = 0.5$. For  models
with  $M \la 0.2~{\rm M}_{\odot}$ we had to increase this temperature to
$T_{\rm eff}=10000$~K
because  models on the lower part of the RGB have temperatures
still too close to $T_{\rm eff}=5000$~K. 

The critical envelope mass, $M_{\rm env}^{\rm crit}$, which marks
the transition between expansion and contraction of the
star's envelope, depends on the mass of the
the hydrogen-exhausted core, $M_{\rm c}$, in a way similar to what is found 
for post-AGB remnants: the larger the core, the smaller the residual envelope
mass. 
When this value is reached, Roche-lobe contact will be shut off, and the 
remnant will continue its evolution with
$L\approx$ $L_{\rm Hyd}$\ $\approx {\rm const}$ and $\dot{R}< 0$ towards 
larger $T_{\rm eff}$. 
The models' evolutionary speed is
determined by the evolution of the envelope mass which is reduced both 
by hydrogen burning, i.e. the core growth rate $\dot{M}_{\rm c}$, and the 
mass loss rate, i.e. $\dot{M}_{\rm R}$ in the presented models. 
Since $\dot{M}_{\rm c}/\dot{M}_{\rm R}\sim 1/R$ 
mass loss becomes rapidly unimportant at least for $T_{\rm eff}> 10000$\,K.
The envelope mass at the turn-around point at maximum effective temperature
is independent of the post-RGB mass loss.

Our procedure is adequate for obtaining reliable
starting models for He-WD tracks.
Due to their evolutionary history
the internal structure of these initial
models is consistent with the one expected from 
binary evolution (see references in Sect. \ref{intro}).
This concerns the envelope masses as well. 
As will be shown in the following sections the latter
has consequences for the mass-radius relation.
Furthermore, the evolutionary envelope masses
give rise to ongoing hydrogen shell burning during
the white dwarf evolution. This greatly
prolongs the cooling times of the He-WDs.

%%
%%
%% Tabelle mit Huellenmasse am RGB (T_eff= 5000 K)
%%
%%
\begin{table}[h]
\caption[]
{\label{krithuell} Total remnant mass, $ M$,
mass of the hydrogen-exhausted core, $M_{\rm c}$,
total mass of the outer hydrogen layers (``thickness''), $M_{\rm H}$,
envelope mass, $M_{\rm env}$, and helium surface
abundance by mass fraction, $Y$,
at $T_{\rm eff}=5000$~K for $M>0.2 {\rm M}_{\odot}$ and at 10000 K
for $M\le 0.2 {\rm M}_{\odot}$ after the end of RGB evolution}
\begin{center}
\begin{tabular}{ccrrc}\hline
      \noalign{\smallskip}
$M/{\rm M}_{\odot}$&$M_{\rm c}/{\rm M}_{\odot}$& $\frac{M_{\rm
H}}{10^{-3} {\rm M}_{\odot}}$&$\frac{M_{\rm env}}{10^{-3}
{\rm M}_{\odot}}$& $Y$ 
\\ 
   \noalign{\smallskip}
\hline
0.179&0.1552 &25.550 &48.289 & 0.464\\
0.195&0.1782 &16.766 &30.768 & 0.462\\ 
0.234&0.2220 &8.118  &13.098 & 0.354\\
0.259&0.2524 &4.771  &7.232 & 0.312\\
0.300&0.2960 &3.189  &4.746 & 0.301\\  
0.331&0.3281 &2.509  &3.744 & 0.301\\ 
0.414&0.4116 &1.446  &2.175 & 0.301\\
\hline 
\end{tabular}
\end{center}
\end{table}

\section{Results}    \label{Sres}

We calculated seven evolutionary sequences starting at different locations along
the RGB of the $ 1~{\rm M}_{\odot}$ sequence. They are listed in 
Table \ref{krithuell}, where we give
the total amount of hydrogen contained in the envelope, the 'thickness'
$M_{\rm H}$, the total envelope mass, $M_{\rm env}$, 
and the helium surface abundance by mass fraction, $Y$. 
Note, that $M_{\rm H}$\ $\approx$ 0.7 $M_{\rm env}$
except for the least massive remnants.
For $M\la 0.3 {\rm M}_{\odot}$ mass loss has uncovered layers where hydrogen
burning took place during the main sequence phase. Consequently, the helium
surface abundance of those models amounts up to $Y\approx 0.46$.
$M_{\rm env}$, and 
$M_{\rm H}$, are reduced (and the core mass correspondingly increased) 
by continuing hydrogen burning.
Mass loss does scarcely reduce the total mass (see Sect.\ \ref{moc}).

At $T_{\rm eff} = 5000$\,K and 10000\,K, resp., after the end
of the RGB evolution the envelope masses of the
Pop.\ I He-WD models are quite large,
varying between $2\cdot 10^{-3}{\rm M}_{\odot}$ 
and $50\cdot 10^{-3}{\rm M}_{\odot}$ (see Table \ref{krithuell}).
It's noteworthy that 
$M_{\rm env}$ would be even larger for lower metallicity
(Castellani et al. 1994). 

\begin{figure*}[th]  
\epsfxsize=18.8cm
\epsfysize=15cm
\epsfbox{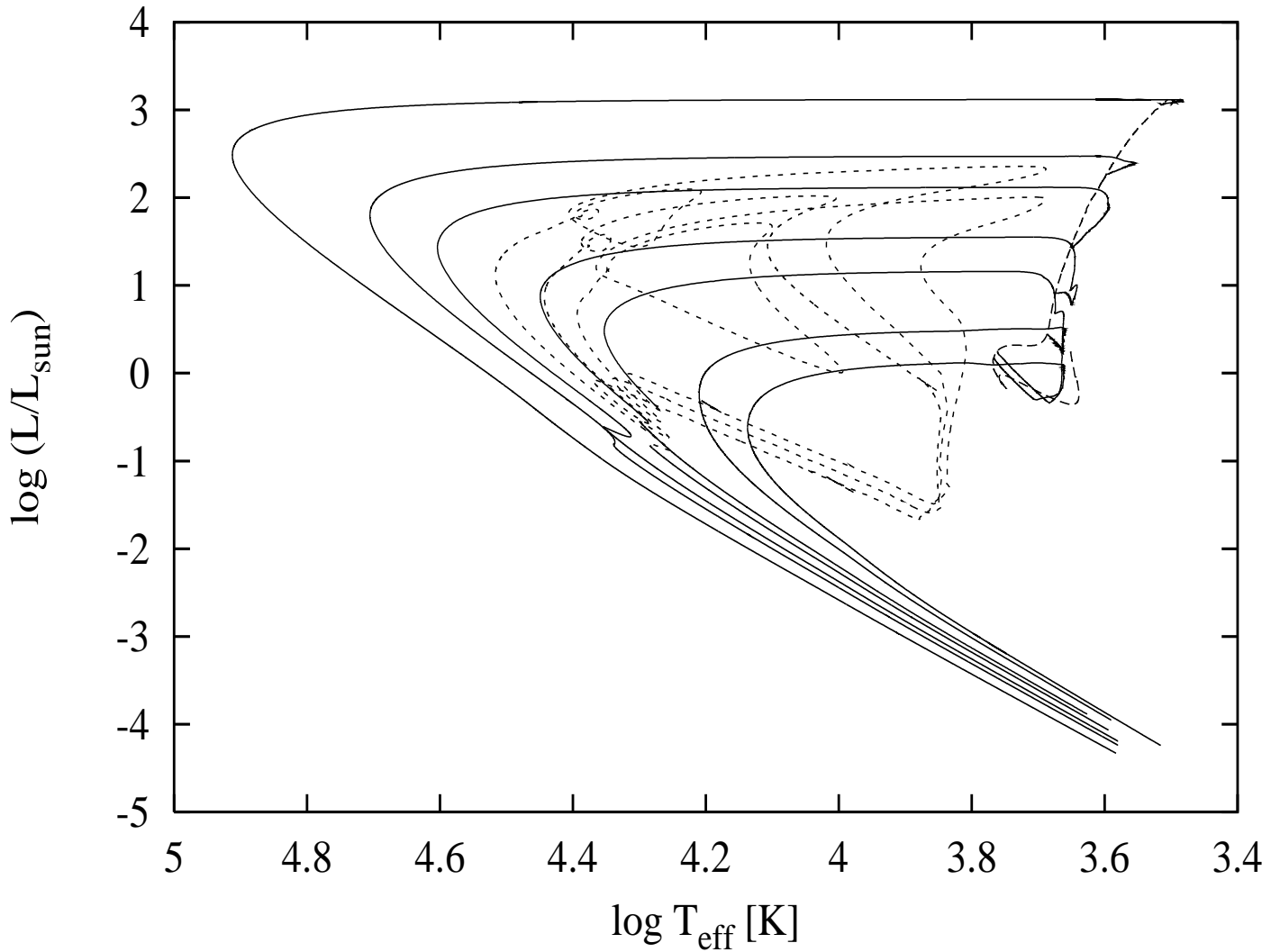}
\caption{ \label{fig2}
HRD with complete evolutionary 
tracks of RGB remnants with different masses 
(from top: 0.414, 0.331, 0.300, 0.259, 0.234, 
0.195, 0.179$\,{\rm M}_{\odot}$). 
The long-dashed curve shows the
evolutionary track of the 1$\,{\rm M}_{\odot}$ star we used for
abstracting the remnants by mass loss (see Sect.~\ref{moc}). 
The short-dashed loops mark the redward short excursions due to hydrogen  
shell flashes that occurred for the 0.259 and 0.234 ${\rm M}_{\odot}$ sequences when 
the CNO cycle is shut off. 
}
\end{figure*}

Figure~\ref{fig2} shows the complete evolutionary tracks of all
sequences down to about $T_{\rm eff}\approx
4000$ K and $ L \approx 10^{-4}$~L$_{\odot}$.
The corresponding data\footnote{Tables 5-11 are only available 
in electronic form at the CDS via ftp 130.79.128.5.}
are given in Table 5-11.
Only for two model sequences, 0.234 $\,{\rm M}_{\odot}$ and
0.259 $\,{\rm M}_{\odot}$, we found
thermal instabilities of the hydrogen-burning shell when CNO burning
ceases (e.g.\
Kippenhahn et al. \cite{KTW}; Iben \& Tutukov \cite{IT86}).
The  hook-like inversions on the cooling tracks
for $ M=0.300 $ and $M=0.331~{\rm M}_{\odot}$
are also due to the onset of unstable burning. Full-scale instabilities are, 
however, avoided because the shell regions cool off too effectively. 
A detailed discussion of the properties of these hydrogen flashes is deferred
to a separate paper. Here we only note that we adjusted our numerical time
steps properly in order to
resolve these flash phases with satisfactory accuracy. 
%
%
% Figure 1: complete HRD 
%
%
It is interesting to remark 
that no instabilities occurred below $M \la 0.20\,{\rm M}_{\odot}$. The existence of
a lower limit for the occurrence of hydrogen flashes, which 
has been predicted by Webbink (\cite{W75}), has important ramifications for any
discussions of the evolution of low-mass white dwarf companions of 
millisecond pulsars, as will be shown below.

Before discussing the properties of our helium white dwarf models,  
the influence of different input physics should  briefly be reported.
Concerning treatment of nuclear reactions we found no significant
differences between tracks 
(or mass-radius relations) using the nuclear network
and those only calculated with equilibrium rates for
the four most important elements (H, He, C, O). 

The comparison between  tracks calculated with 
OPAL  and  the older CS
opacities yielded, in short, the following differences:
\begin{itemize} 
\item A well known shift of the RGB locus towards higher effective  
       temperatures for CS opacities. 
\item  Slightly larger turn-around temperatures for the 
       CS post-RGB tracks. For example,  a CS model sequence 
      with $M=0.412~{\rm M}_{\odot}$ reached  $\log T_{\rm eff} = 4.915$, the corresponding 
      OPAL model ($M=0.414~{\rm M}_{\odot}$) only $\log T_{\rm eff}=4.911$. 
      %%% (no nuclear network) and $T_{\rm eff}=4.912$ for $M=0.414 \,{\rm M}_{\odot}$ (with
      %%%network), resp. {\tt hier nur OPAz vergleichen Was meint das ???}
\item The opacity effects  on the locations of
      the cooling tracks as well as on the mass-radius relation
      and envelope masses  were found
      %%%'thickness' of the outer hydrogen layer were found
      to be small. Accordingly, the opacities seem not to be
      a critical parameter for the determination of the mass-radius relation
     (for given metallicity).
\item White dwarf cooling ages are, as expected, significantly
      effected by the opacities. Models calculated with
      CS opacities have shorter cooling ages than those calculated
      with OPAL. 
      The influence, however, is only noticeable in our more massive
      models where hydrogen burning is less dominant (see next section).
\end{itemize}
\subsection{White dwarf cooling properties}   \label{SScool}
For CO white dwarfs it is well known that the energy contribution of 
nuclear burning
drops rapidly below the gravothermal energy release 
(compressional heating plus cooling)
when the cooling branch is reached (Iben \& Tutukov \cite{IT84b};
Koester \& Sch\"onberner \cite{KS}; Bl\"ocker \cite{Bb}). 
In contrast, full evolutionary calculations for helium white dwarfs show
that hydrogen shell burning still plays a significant role on the
cooling branch leading to hydrogen shell flashes for certain masses
(e.g. Kippenhahn et al.\ 1968, Iben \& Tutukov 1986, Castellani
\& Castellani 1993). The present calculations illustrate that
hydrogen shell burning remains the main energy source down 
to temperatures well below $T_{\rm eff}=10\,000$~K on the cooling branch.
For demonstration, the temperatures in the burning shells  
are listed in Table \ref{tschale}. Also given are the corresponding
envelope masses. The comparison with those at maximum effective temperatures
illustrates clearly the significance of hydrogen burning.

\begin{table*}[htb]
\caption[]
{\label{tschale} Temperatures at the center and the lower and 
upper boundary of the hydrogen burning shell for our
different white dwarf models at $T_{\rm eff}=10\,000$ K on the cooling branch. 
$T_{\rm bot}$
and $T_{\rm top}$ are the temperatures (in K) of those layers
where the energy generation rate for hydrogen burning,
$\epsilon_{\rm H}$, has dropped to 1\,\% of the maximum value.
${\rm M}_{\rm env,knee}$ and ${\rm M}_{\rm env}$ are the envelope masses
at maximum effective temperature ($T_{\rm eff,knee}$) and at
$T_{\rm eff}=10\,000$~K on the cooling branch, resp.
$L_{\rm knee}$ is the
surface luminosity corresponding to $T_{\rm eff}=T_{\rm eff,knee}$.
}
\begin{center}
\begin{tabular}{cccccccc}\hline
   \noalign{\smallskip}
$M/{\rm M}_{\odot}$&$\log (T_{\rm c}/{\rm K})$&$\log (T_{\rm bot}/{\rm K})$&
$\log (T_{\rm top}/{\rm K})$&$\frac{{\rm M}_{\rm env}}{10^{-3}{\rm M}_{\odot}}
$& $\frac{{\rm M}_{\rm env,knee}}{10^{-3}{\rm M}_{\odot}}$&
$\log T_{\rm eff,knee}/{\rm K}$&
$\log (L_{\rm knee}/L_{\odot})$ \\ 
\noalign{\smallskip}
\hline
0.179 &7.092 & 7.086 & 6.796 &1.460 &5.369 & 4.1372 & -0.6202\\
0.195 &7.070 & 7.068 & 6.784 &1.184 &5.214 & 4.2100 & -0.2262\\
0.234 &7.039 & 7.029 & 6.766 &0.744 &2.836 & 4.4430 &  0.8708\\
0.259 &7.022 & 7.011 & 6.748 &0.666 &2.593 & 4.5168 &  1.1283\\
0.300 &7.000 & 6.988 & 6.730 &0.529 &1.652 & 4.6038 &  1.4364\\
0.331 &6.987 & 6.972 & 6.721 &0.438 &1.216 & 4.7053 &  1.7950\\
0.414 &6.959 & 6.934 & 6.704 &0.317 &0.612 & 4.9124 &  2.4873\\ \hline 
\end{tabular}
\end{center}
\end{table*}

\begin{figure}[th]
\epsfxsize=8.8cm
\epsfbox{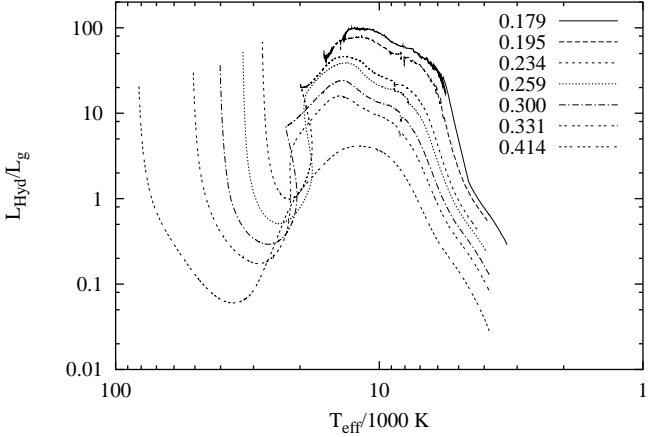}
\caption{\label{fig5} Ratio of hydrogen to gravothermal luminosity
as a function of $T_{\rm eff}$ for He-WDs of different masses. }
\end{figure}

\begin{figure}[h]
\epsfxsize=8.8cm 
\epsfbox{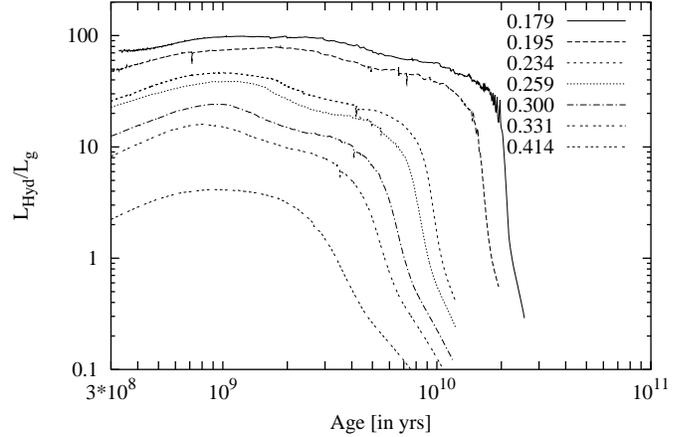}
\caption{\label{fig6} Ratio of hydrogen to gravothermal luminosity
as a function of the cooling age $\tau$. Only the lower parts
of the respective cooling curves ($T_{\rm eff} < 15000\dots 18000$~K) are plotted.
The ages are counted from $T_{\rm eff} = 10\,000$~K ($M<0.2~{\rm M}_{\odot}$) and
$T_{\rm eff} = 5\,000$~K ($M>0.2~{\rm M}_{\odot}$)
on the (horizontal) post-RGB branch shown in Fig.~1. } 
\end{figure}

Figure~\ref{fig5} shows the ratio of the hydrogen shell luminosity, 
$L_{\rm Hyd}$, 
 to the gravothermal luminosity, $L_{\rm g}$,  for our helium-white dwarf
sequences as a function of $T_{\rm eff}$. The shell flashes are omitted for
clarity. Figure~\ref{fig6} shows the same ratio
as a function of the cooling age $\tau$.
%
%
%
%
% Figure 2: L_hyd/L_G as Fct. of T_eff
%
%
%
%
%
%
% Figure 3: L_hyd/L_G als Fkt. von tau
%
%

The left part of Fig.~\ref{fig5} indicates a rapid
drop of $L_{\rm Hyd}/L_{\rm g}$\ due to the decline of CNO burning right
 after the models have entered their cooling branches.
At lower luminosities and temperatures,
$L_{\rm Hyd}/L_{\rm g}$\ exceeds unity again, entirely due to pp burning
(see below). This phase of dominant hydrogen burning lasts for about 20~Gyr
for the two least massive models (see Fig.~\ref{fig6}).
Even the  0.414 ${\rm M}_{\odot}$\ model burns hydrogen for
about 3~Gyr until gravothermal energy release finally resumes.

\begin{figure}[htb]
\epsfxsize=8.8cm 
\epsfbox{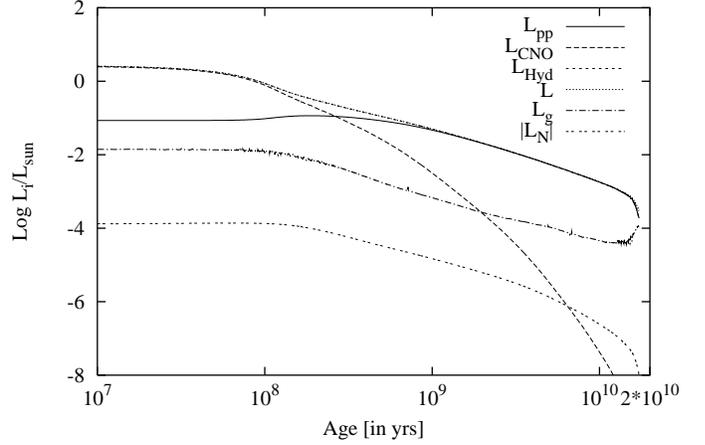}
\caption{\label{picl1} Different luminosity contributions $L_i$
as a function of cooling time $\tau$ for the 0.195 $\,{\rm M}_{\odot}$-sequence. 
Ages are counted from $T_{\rm eff} = 10\,000$~K on the horizontal post-RGB branch.
}
\end{figure}

\begin{figure}[bht]
\epsfxsize=8.8cm 
\epsfbox{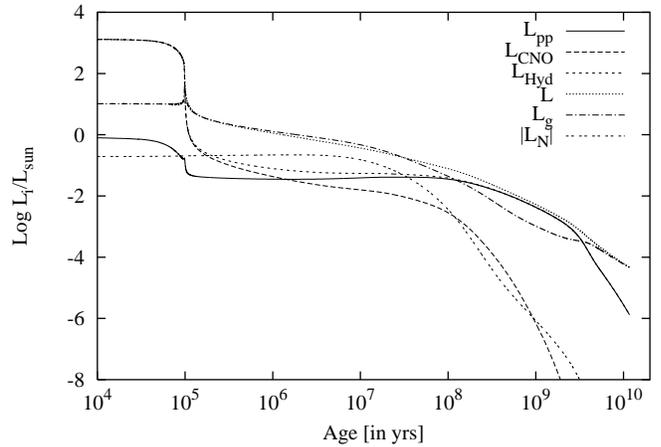}
\caption{\label{picl2} 
Same as Fig.~ \ref{picl2},
but for the 0.414$\,{\rm M}_{\odot}$-sequence. Ages are counted from 
$T_{\rm eff} = 5\,000$~K on the horizontal post-RGB branch.}
\end{figure}

A more detailed picture of the temporal evolution of the  
different luminosity
 contributions (Figs.~\ref{picl1} and
\ref{picl2}) reveals that at higher luminosities
almost the entire energy production comes from shell
CNO burning. At about $\tau\approx
2\cdot10^8$~yr ($0.195~{\rm M}_{\odot}$) and $\tau\approx 1\cdot10^5$~yr 
($0.414~{\rm M}_{\odot}$) the cooling branch is reached, and
$L$, $L_{\rm Hyd}$\ and 
$L_{\rm CNO}$ start to  decrease, slowly in the low-mass model, but very 
rapidly in the $0.414~{\rm M}_{\odot}$\ model. In the less massive models hydrogen
burning via the pp chains remains the predominant luminosity contribution 
for more than 10~Gyr until finally gravothermal energy release starts to take
over (see Fig.~\ref{picl1}). 
Since unstable burning is restricted to a
certain mass range the feature of continuing pp burning of low mass He-WDs
($M\la 0.2 {\rm M}_\odot$) is particularly independent of the details in
modelling the phase of hydrogen shell flashes (e.g. the mass loss during
the reexpansion phase due to Roche lobe overflow).
For more massive
models, pp burning stays important only for a few Gyr (Fig.~\ref{picl2}).

The more gradual transition from CNO burning to pp burning is typical
for less massive He-WDs.  
The more massive models show a rapid, more step-like shaped transition,
very similar to the situation found in CO-WDs (Bl\"ocker 
\cite{Bb}).
Figures~\ref{picl1} and \ref{picl2}, together with Fig.~\ref{fig5}, also 
demonstrate that neither the gravothermal contribution nor the energy
loss via neutrinos plays a significant role in the
cooling history of He-WDs with $M \la 0.2 \,{\rm M}_{\odot}$ for
ages $\tau \la 10$ Gyr. 

\begin{figure}[h]
\epsfxsize=8.8cm 
\epsfbox{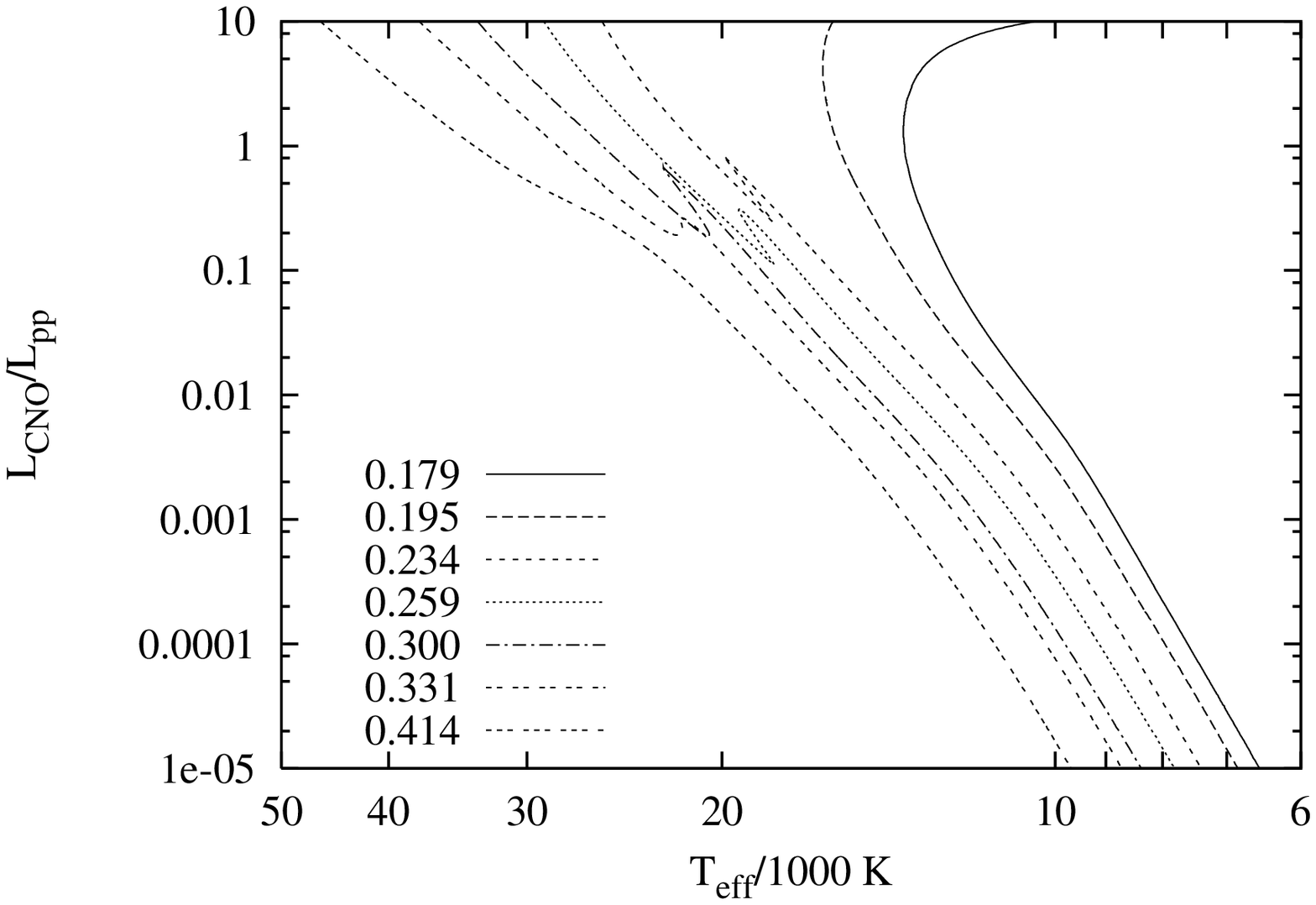}
\caption{\label{fig6a} Contribution of the two
hydrogen burning reaction chains: ratio of luminosity
due to the pp-chain ($L_{\rm pp}$) and due to the
CNO-cycle ($L_{\rm CNO}$) as a function of $T_{\rm eff}$ on
the WD cooling branch. The shell flashes are omitted for clarity.
}
\end{figure}

Figure~\ref{fig6a} illustrates how the 
the ratio $L_{\rm CNO}/L_{\rm pp}$\ varies during the course of evolution.
The hooks indicate the onset of unstable CNO burning, the flash loops 
are omitted for clarity. 
Figure \ref{fig6a} elucidates also that
pp burning becomes the dominant energy contribution
for He-WDs below $T_{\rm eff}\approx 35\,000$~K for
$M=0.414\,{\rm M}_{\odot}$ and below $T_{\rm eff}\approx 18\,000$~K for
$M=0.234 \,{\rm M}_{\odot}$. This coincides with the change from CNO to
pp burning as can be seen 
from the minima in Fig.~\ref{fig5}.

Summarizing, nuclear burning remains an important, if not dominant, energy
source for helium-white dwarfs, even for temperatures of $T_{\rm eff}=10\,000$~K 
and below. This is in particular true for objects with $M< 0.2 \,{\rm M}_{\odot}$,
like the companion of \object{PSR J1012+5307}. For example, Fig.~\ref{fig5}
shows that for the WD companion of \object{PSR J1012+5307} 
with $M\approx 0.2 \,{\rm M}_{\odot}$ the ratio $L_{\rm Hyd}/L_{\rm g}$\
is at least 50. The assumption $L_{\rm nuclear}\approx 0$ usually 
made in contraction models thus appears not to be appropriate for He-WDs. 
(e.g.\ Chin \& Stothers \cite{ChSt}; Althaus \& Benvenuto \cite{AB}).

However, Hansen \& Phinney (\cite{HPa}) and Benvenuto \& Althaus (\cite{BA}) 
did consider nuclear burning but found it
of little importance. They did not model the evolution
of the He-WD progenitor and had to make 
assumptions about the hydrogen
envelope mass.
Hansen \& Phinney (\cite{HPa}) selected an envelope mass of 
$M_{\rm H}=3\cdot10^{-4}~{\rm M}_{\odot}$
taken at some stage during the end of the second hydrogen shell flash from
a $0.3~{\rm M}_{\odot}$ model sequence calculated by Iben \& Tutukov (\cite{IT86}).
At this phase the envelope mass has already been
considerably reduced by the hydrogen burning. Additionally,
the Iben \& Tutukov (1986) model had suffered from severe mass loss
due to Roche lobe overflow during the expansion back to the RGB domain.
Note that in models with $M\approx 0.3 {\rm M}_{\odot}$
and such a small hydrogen envelope mass nuclear 
burning indeed becomes less important.
However, as mentioned above, shell flashes do \emph{not} occur 
for $M \la 0.20\,{\rm M}_{\odot}$ which is the relevant mass range 
for the \object{PSR J1012+5307}\ companion. 
The evolutionary models
show that this object should have
a hydrogen envelope of  $M_{\rm H}\approx 10^{-3}~{\rm M}_{\odot}$.
Thus, the finding by Hansen \& Phinney (\cite{HPa})
that nuclear burning is negligible
results from their choice of the hydrogen envelope mass which
is not appropiate according to our models. 

Additionally, the evolutionary history plays an important role as well
since it determines the thermomechanical structure on the
prevailing part of the cooling branch which is different 
if the pre-He-WD evolution is not considered.
For instance, in contrast to full evolutionary calculations 
(e.g. Iben \& Tutukov 1986, present paper)
hydrogen burning can be found to be unimportant even
for thick-envelope models ($M_{\rm H}\approx 10^{-3}~{\rm M}_{\odot}$
for $M=0.2\,{\rm M}_\odot$)
if the pre-He-WD evolution is not considered
(Benvenuto \& Althaus \cite{BA}). 

%
%
%
% Figure 4: luminosity evol. 0.195
%
%
%
%
%
%
%
% Figure 5: luminosity evol. 0.414
%
%
%
%
%%\subsection{Contraction models}  \label{SScontrac}
%%
%
\subsection{Age and mass of the PSR J1012+5307 white dwarf companion}
In order to estimate the influence of initial conditions on age and mass
determinations for low-mass white dwarfs, 
we also calculated contraction models ($L_\mathrm{nuc}=0$)
with identical masses and chemical profiles as our evolutionary models.
For that purpose we started with a
homogeneous main sequence model of a given mass whose chemical
profile was adapted over a series of about 50 models 
to the chemical structure of the corresponding evolutionary model
at the beginning of the cooling branch (see Bl\"ocker et al. \cite{BHDB}), 
but did not allow for hydrogen burning.  

%
%
%
% Figure 6: Hydrogen burning: L_CNO/L_pp (T_eff)
%
%

\subsubsection{Cooling behaviour of helium-white dwarfs} 
%
%
%
%
% Figure 7: Log g -T_eff mit 0.195 M_sol
%
%
%
%
\begin{figure*}[th]
\epsfxsize=17cm 
\epsfbox{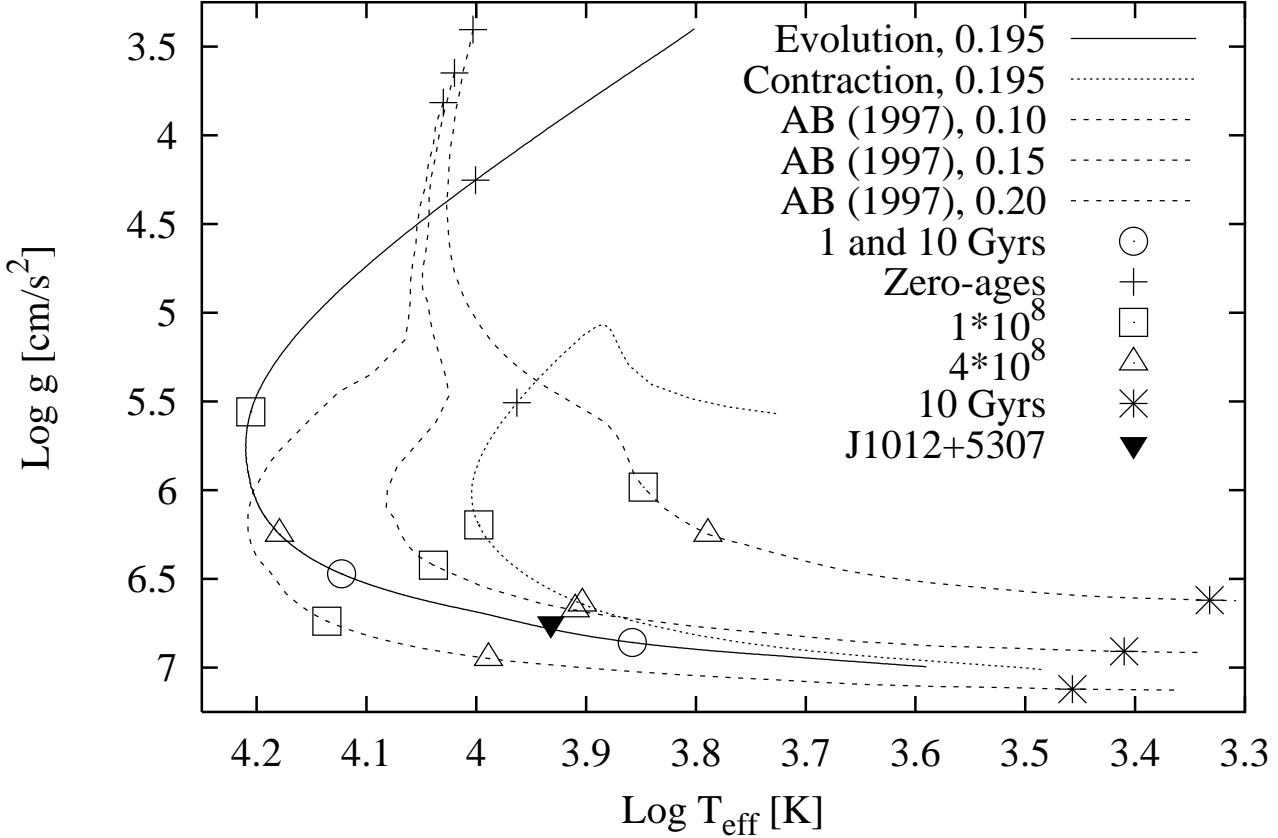}
\caption{\label{fig7a} $\log g-\log T_{\rm eff}$ diagram with our
tracks for $M=0.195 \,{\rm M}_{\odot}$ with and without consideration
of nuclear burning (continuous and dotted line, see text). 
Also shown are the tracks of Althaus \& Benvenuto (\cite{AB}) 
for $M=0.1, 0.15$ and $0.2 \,{\rm M}_{\odot}$ (AB, dashed lines) 
as well as time marks for different cooling stages as indicated. 
Additional circles mark cooling ages of our evolutionary model of 1 and
10~Gyr, resp. All ages are counted from a pre-white dwarf stage close to 
$T_{\rm eff}=10\,000$~K.
The filled triangle marks the position of the PSR
J1012+5307 white dwarf. }
\end{figure*}
Figure~\ref{fig7a} illustrates in a $\log g,T_{\rm eff}$ diagram the differences in the
cooling properties between contraction models and one of our evolutionary 
models  (0.195~${\rm M}_{\odot}$). At the position of the \object{PSR J1012+5307}
companion, our evolutionary model indicates an age of about $6 \pm 1
\mathrm{Gyr}$.  In contrast, our contraction model as well as the models of
Althaus \& Benvenuto (\cite{AB}) give only
a cooling age of 0.4~Gyr,   about a factor
of 15 smaller and in serious conflict with the pulsar's spin-down age.
Note that the contraction models agree in the cooling age but differ 
considerably in the mass-radius-relation (see Sect.\ \ref{MRR}),  
since Althaus \& Benvenuto (\cite{AB}) used $M_{\rm H}=0$
while we took the same $M_{\rm H}$ as in the corresponding
evolutionary model.

%  
% Figure 8: Log g - log T_eff with isochrones  (Pulsar-Alter)
% 
% 
\begin{figure}[h] 
\epsfxsize=8.8cm  
\epsfbox{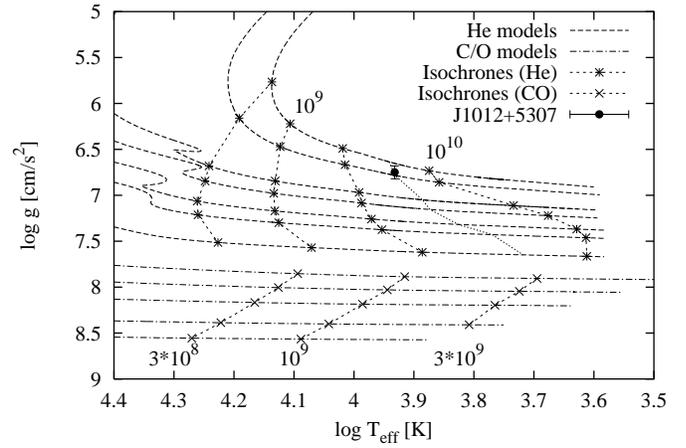}
\caption{\label{piciso}
$\log g-\log T_{\rm eff}$ diagram with our evolutionary tracks
for  He-WDs  ($M = 0.179$, 0.195, 0.234, 0.259, 0.300, 0.331, 
$0.414 {\rm M}_{\odot}$ and for CO-WDs ($M = 0.524$, 0.605, 0.696, 0.836, 
$0.940 {\rm M}_{\odot}$; Bl\"ocker 1995b). The masses increase from top to 
bottom. Also shown is the locus of the
companion of the millisecond pulsar PSR J1012+5307 as given by van Kerkwijk
et al. (1996).  Isochrones are shown for   $3\cdot 10^8, 1\cdot 10^9,
3\cdot 10^9 ,{\rm and} \, 10^{10} \,{\rm
yr}$ (from left to right, last isochrone only for  He-WDs). 
The isochrones for the He-WDs are turned over and shifted
to the left because hydrogen burning
slows down the evolution so much. The dotted isochrone of $6\cdot
10^9$ yr fits the position of the PSR J1012+5307 companion.}
\end{figure} 
A more detailed view of the cooling behaviour of our evolutionary 
models is given in  Fig.~\ref{piciso}
which shows the tracks as well as the corresponding isochrones
in the  $\log g$ -- $\log T_{\rm eff}$ plane. 
Note that hydrogen burning leads to different slopes
of the isochrones for CO- and He-WDs.
Usually, more massive white dwarfs cool faster then less massive ones.
For He-WDs, however, the reverse is true. Due to the increasing
importance of nuclear burning with decreasing mass, the less massive ones
are considerably older than the more massive ones, at a given temperature
below about 10\,000~K. Hence,
the isochrones of He-WDs run almost perpendicular to those of
CO-WDs, and an extrapolation of cooling times of CO-WDs
into the He-WD regime would give significantly lower cooling ages, close to
those of the contraction models.
  
As already mentioned above, our evolutionary models predict a
cooling age of $6\pm 1 \mathrm{Gyr}$ for the white dwarf companion
of \object{PSR J1012+5307}, which is in good agreement with the
estimated pulsar's spin-down age of 7 Gyr, and also with age estimates 
of Alberts et al. (\cite{ASH}) who also included nuclear burning.

Thus, evolutionary models for low-mass white dwarfs which start from explicit 
model calculations of the progenitors evolution
seem to indicate that neither a modification of the theory of field decay as
proposed by Burderi et al. (\cite{BKW}) nor the consideration of initial spin periods
close to the present one are necessary to resolve the apparent age
discrepancy between the two components of \object{PSR J1012+5307}. 
\subsubsection{Mass-radius relations of helium-white dwarfs}\label{MRR}
\begin{figure}[th] 
\epsfxsize=8.8cm  
\epsfbox{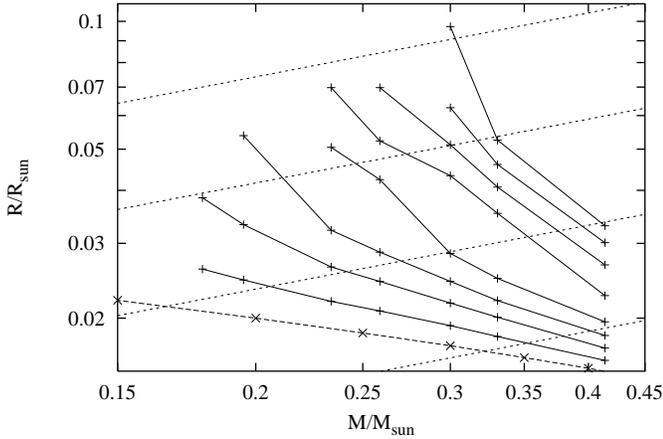}
\caption{\label{mrrpic} 
Mass-radius relation for He-WDs at different
effective temperatures,  calculated from our evolutionary models 
from $T_{\rm eff}=40000$ K (upper line) down to $T_{\rm eff}=5000$ K
(lower line) in steps of $\Delta T_{\rm eff}=5000$ K (Table \ref{mrrtab}).
For comparison also the $T=0$ K relation of Hamada \& Salpeter (1961) for pure
helium models (dashed line) is given. Along the dotted lines gravity is
constant, with (from above) $\log g= 6.0, 6.5, 7.0$ and 7.5).
Temperatures larger than 10\,000~K are not reached by RGB remnants of too low
a mass (see Fig.~\ref{fig2}).
The irregularities around $T_{\rm eff}=20000$ K
are due to thermal instabilities in
the hydrogen burning shell as explained in the text.  }
\end{figure}

The mass-radius relation according to our models is
shown in Fig.~\ref{mrrpic} and also listed in Table \ref{mrrtab}.
For intermediate temperatures some distortions can be seen where 
hydrogen flashes with the associated high burning rates lead to a
substantial reduction of the envelope mass. 
A more thorough discussion will be presented in a forthcoming paper.
Important for the interpretation of observations is the remarkable
temperature dependence of the radii even at the low temperature end.
%
%
%
%  
% Tabelle mit MRR, neue Version (2-spaltig)
%  
%  
%
\addtolength{\tabcolsep}{2mm}
\renewcommand{\arraystretch}{1.2}
\begin{table*}[htb]
\caption[]
{\label{mrrtab} Radii for He-WDs for different masses and
effective temperatures (see also Fig.~ \ref{mrrpic}).
The radii are given in solar units.}
\begin{center}
\begin{tabular}{rccccccc}
   \cline{2-8}
\multicolumn{8}{c}{$M/{\rm M}_{\odot}$}\\ \cline{2-8}
& 0.179 & 0.195 & 0.234 & 0.259 & 0.300 & 0.331 & 0.414\\ \hline
$T_{\rm eff}/\mathrm{K}$ & & & & & & & \\ \hline
40\,000 & - & - & - & - & 0.0972 & 0.0525 & 0.0330\\ 
35\,000 & - & - & - & - & 0.0626 & 0.0460 & 0.0301\\ 
30\,000 & - & - & - & 0.0698 & 0.0512 & 0.0407 & 0.0267\\ 
25\,000 & - & - & 0.0698 & 0.0523 & 0.0433 & 0.0353 & 0.0226\\ 
20\,000 & - & - & 0.0505 & 0.0424 & 0.0284 & 0.0248 & 0.0196\\ 
15\,000 & - & 0.0538 & 0.0322 & 0.0286 & 0.0244 & 0.0220 & 0.0182\\ 
10\,000 & 0.0384 & 0.0332 & 0.0264 & 0.0244 & 0.0217 & 0.0201 & 0.0170\\ 
5\,000  & 0.0261 & 0.0246 & 0.0219 & 0.0208 & 0.0192 & 0.0181 & 0.0159\\ 
\hline
\end{tabular}
\end{center}
\end{table*}
%
%
%  Figure 9 - MRR
%
% 
%
%
%
%
%
% Figure 10: Log g - Log T_eff mit Vergleich EVOL/KONT
%
%
\begin{figure}[h]
\epsfxsize=8.8cm 
\epsfbox{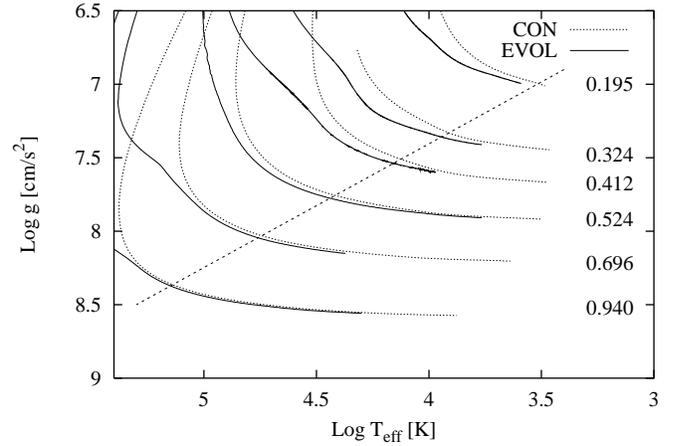}
\caption{\label{fig7b}
$\log g-\log T_{\rm eff}$-diagram with WD tracks
according to evolutionary models (i.e.\ with consideration
of nuclear burning, solid curves) and contraction models (i.e.\ no
nuclear burning, dotted curves) with the same masses and chemical structures
for the masses as indicated by the labels.
The evolutionary tracks are either from this study or
from  Bl\"ocker (\cite{Bb}) for the larger masses. The contraction models are
from Bl\"ocker et al. (\cite{BHDB}).
The dashed line divides the diagram roughly in two parts: 
in the lower right part contraction and evolutionary models
have virtually the same thermo-mechanical structure. }
\end{figure}
%
%
%
%
%
%
% Figure 11: MRR mit Vergleich EVOL/KONT
%
%
\begin{figure}[h]
\epsfxsize=8.8cm 
\epsfbox{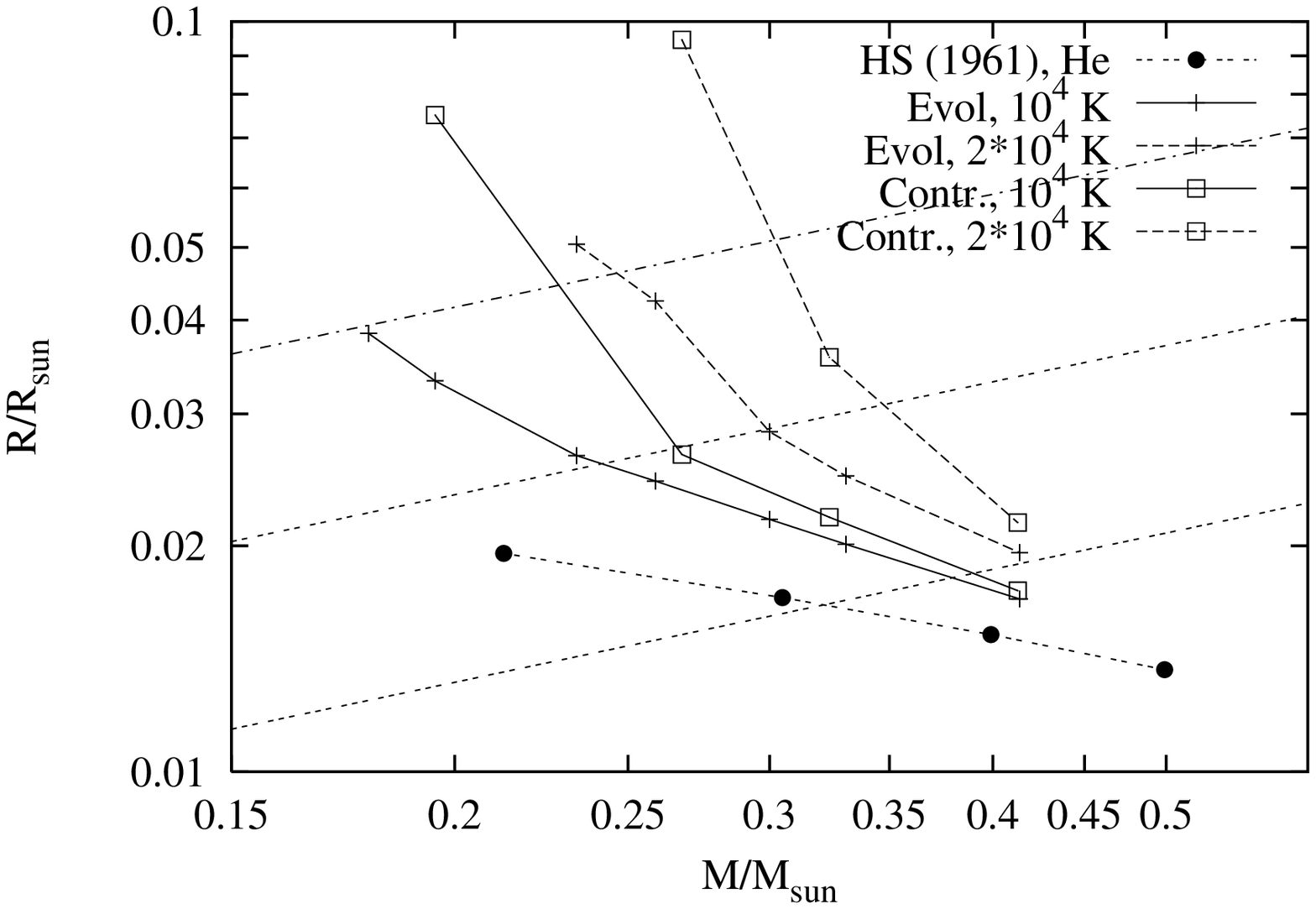}
\caption{\label{fig7c} Mass-radius relation derived from
evolutionary (crosses) and contraction models (squares)
with identical chemical profiles 
at $T_{\rm eff}=10\,000$ K and 20\,000~K.
Also given are the $T=0$ K relation of 
Hamada \& Salpeter (\cite{HS}) (dots) and lines of constant surface 
gravities (dashed, from above: $\log g = 6.5, 7.0, 7.5$).}
\end{figure}

The structural
differences between evolutionary and purely contracting white dwarf models
influence the evolutionary tracks as well.
For selected masses the tracks of evolutionary and contracting white dwarfs with
identical chemical structures are plotted in Fig.~\ref{fig7b}.
The long-dashed line in the figure connects roughly the loci where the
tracks of evolutionary and contraction models do converge. In the regime
above this line the thermo-mechanical structure of white dwarfs does depend
on their (assumed) evolutionary history. 
This has important consequences for mass determinations from spectroscopy
(see Bl\"ocker et al. \cite{BHDB}). 
If an object's position lies above the line 
its mass will be overestimated by using simple contraction models
with the same envelope masses as the corresponding evolurionary models. This fact
is especially important for helium-white dwarfs with their low masses: 
The tracks of evolutionary and contraction models in the
$\log g-\log T_{\rm eff}$-diagram merge only very late at rather low
effective temperatures. For example, the merging of both 
0.2 $\,{\rm M}_{\odot}$ tracks 
occurs at $\log T_{\rm eff} \approx 3.5$\,!

Although in the lower right part of Fig.~\ref{fig7b} the 
thermo-mechanical structures of evolutionary and contracting models agree, the
evolutionary speeds may be very different because  hydrogen burning slows
down the 'cooling' for the lower masses.
    
\begin{figure}[h] 
\epsfxsize=8.8cm  
\epsfbox{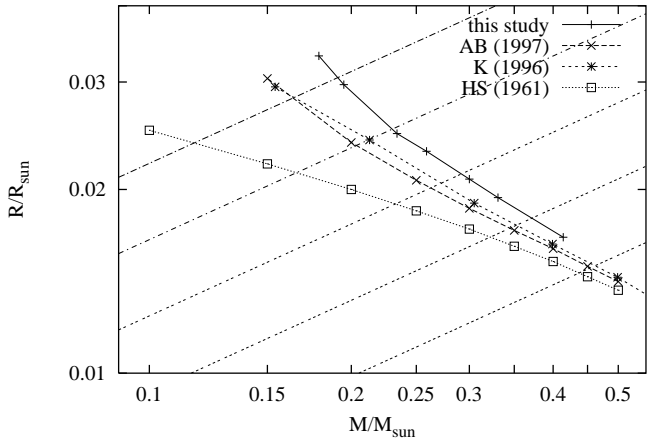}
\caption{\label{pulsmrr} Mass-radius relation at $T_{\rm eff}=8550$ K, relevant
for the mass determination of the  \object{PSR J1012+5307}) companion, from 
different authors: AB = Althaus \& Benvenuto (\cite{AB}), 
K = van Kerkwijk et al. (\cite{KBK}); HS = Hamada \& Salpeter (\cite{HS}).
Lines for constant surface gravity are also given, starting from $\log g =6.75$
(upper line, relevant for \object{PSR J1012+5307}) to $\log g=7.75$ in 
steps of $\Delta \log g=0.25$.
} 
\end{figure}

The possible differences in structure between evolutionary and contracting
white dwarf models translate also into differences in the  mass-radius relations,
as demonstrated in  Fig.~\ref{fig7c}. One can again see that according to
contraction models the object's mass for given gravity is overestimated.
However, the larger hydrogen envelope masses which result in our
models from their evolutionary history have the effect to 
increase the masses derived from atmospheric parameters and mass-radius
relation. Thus the two effects of contraction vs.\ evolutionary
models (with regard to the treatment of $L_\mathrm{nuc}=0$)
and smaller, mass independent envelope masses vs.\ evolutionary
(thicker, mass dependent) envelope masses compensate
each other to a certain extent. This can be seen from
Fig.~\ref{pulsmrr} which gives mass-radius relations of different
origin in order to compare the mass determination of the
\object{PSR J1012+5307}\ companion. The deviations from the 
$T=0$ K relation of Hamada \&
Salpeter (\cite{HS})  for a pure helium composition demonstrate the
well known importance of finite temperature effects in the equation of state,
especially for the low-mass WDs. For a surface
gravity $\log g=6.75$ of the pulsar's companion that relation yields
$M\approx 0.12~{\rm M}_{\odot}$ which is close to the lower limit $M \cdot \sin i =
0.11 \,{\rm M}_{\odot}$. 
The differences 
between our mass-radius relation and those
of van Kerkwijk et al. (\cite{KBK}) and Althaus \& Benvenuto (\cite{AB})
are mainly due to the larger envelope masses 
(i.e.\ thicker hydrogen layers) predicted from our evolutionary models. 
For example, van Kerkwijk et al. (\cite{KBK}) interpolated between the Hamada \&
Salpeter relation based on pure helium cores ($M_{\rm H}=0$) and that based on
models from Wood (\cite{Wma}) with 'thick', but constant hydrogen layers  
($M_{\rm H}$\ = $10^{-4}M_\ast$). Our evolutionary calculations, however,
revealed a large dependence of envelope thicknesses
on total  masses: $M_{\rm H}$\ drops from
$\approx 10^{-3} \,{\rm M}_{\odot}$ for $M=0.2~{\rm M}_{\odot}$ to $M_{\rm H}$\ $\approx 
10^{-4}~{\rm M}_{\odot}$ for a C/O WD with $M=0.6~{\rm M}_{\odot}$,
the heaviest He-WDs with $M\approx 0.45\,{\rm M}_{\odot}$ having about 
$M_{\rm H}$\ $=10^{-3.75}~{\rm M}_{\odot}$ (see Table \ref{tschale}, Bl\"ocker et al.\
\cite{BHDB}). Note that in contrast to Figs.\ \ref{fig7b} and \ref{fig7c}
the contraction models of Althaus \& Benvenuto (\cite{AB}) give smaller
white dwarf masses compared to our evolutionary models due to the much smaller
envelope masses (in this case $M_{\rm H}=0)$.
%
%
%%%  Figure  12 - MRR for 8550 K
%
%
%

As can be seen from Fig.~\ref{pulsmrr}, our mass-radius relation yields
a mass of $M=0.19\pm 0.02~{\rm M}_{\odot}$ for the white dwarf companion
of \object{PSR J1012+5307}, using $\log g=6.75$.  %%%%%  Fehler???
This is larger than the value
quoted by van Kerkwijk et al. (\cite{KBK}) and Callanan et al. (\cite{CGK})
but within the mass range
$M_{\rm He-WD}=0.13\dots 0.21\, {\rm M}_{\odot}$ given by Hansen \& Phinney
(\cite{HPb}). Similarly, 
$M=0.16~{\rm M}_{\odot}$ follows from the Althaus \& Benvenuto (\cite{AB})
relation. With the spectroscopic analysis from Callanan et al. (\cite{CGK})
a mass of $M=0.15\pm 0.02\, {\rm M}_{\odot}$ follows by extrapolating
our mass-radius relation.

With the mass ratio $M_\mathrm{pulsar}/M_\mathrm{He-WD} \approx 9.5\pm 0.3$
(van Kerkwijk 1998, priv.\ comm.) the pulsar's mass can be estimated:
$M_{\rm He-WD}=0.15~{\rm M}_{\odot}$ gives 
$M_{\rm pulsar}\approx 1.43\pm 0.25 ~{\rm M}_{\odot}$,
close to the canonical pulsar mass of $\approx 1.5~{\rm M}_{\odot}$.
On the other hand, $M_{\rm He-WD}=0.19~{\rm M}_{\odot}$
gives $M_{\rm pulsar}\approx 1.81\pm 0.25 ~{\rm M}_{\odot}$.
Slightly increased values for $M_{\rm pulsar}$ are obtained if the
mass ratio $M_\mathrm{pulsar}/M_\mathrm{He-WD} \approx 10.5\pm 0.5$
from Callanan et al. (\cite{CGK}) is applied.
A summary of the results for the mass determination of the components
of the \object{PSR J1012+5307} system is given in Table \ref{mastab}.
\begin{table}[h]
\caption[]
{\label{mastab}Spectroscopic results, i.e. $\log g$ and $T_{\rm eff}$,
and derived masses, $M_{\rm He-WD}$,
for the \object{PSR~J1012+5307} He-WD.
The lower part gives the He-WD mass from the present work ($M_{\rm He-WD, PW}$)
derived from the respective spectroscopic analysis.
$M_{\rm pulsar, K}$ is the pulsar mass based on the
mass ratio $M_\mathrm{pulsar}/M_\mathrm{He-WD} \approx 9.5\pm 0.3$
(van Kerkwijk et al.\ 1998, priv.\ comm.) and $M_{\rm He-WD, PW}$.\
$M_{\rm pulsar, C}$ is the pulsar mass based on the mass ratio
from Callanan et al. (\cite{CGK}) ($M_\mathrm{pulsar}/M_\mathrm{He-WD}
\approx 10.5\pm0.5$) and $M_{\rm He-WD, PW}$.}
\begin{center}
\begin{tabular}{p{2.25cm}p{2.0cm}p{2.0cm}}\hline
&Callanan et al. (\cite{CGK})& 
van Kerkwijk et al. (1996)\\ \hline
$\log g [\rm{cm/s^2}] $& $6.34\pm 0.20$& $6.75\pm 0.07$\\
$T_{\rm eff}$ [K]& $8670\pm 300$ & $8550\pm 25$\\
$M_{\rm He-WD} [{\rm M}_{\odot}]$&$0.16\pm 0.02$&$0.16\pm 0.02$\\
\hline\hline
$M_{\rm He-WD, PW} [{\rm M}_{\odot}]$&$0.15\pm 0.02$&$0.19\pm 0.02$\\
$M_{\rm pulsar, K} [{\rm M}_{\odot}]$&$1.43\pm 0.25$&$1.81\pm 0.25$\\
$M_{\rm pulsar, C} [{\rm M}_{\odot}]$&$1.59\pm0.30$&$2.00\pm0.30$\\
\hline
\end{tabular}
\end{center}
\end{table}

\section{Summary and conclusions}  \label{Sconcl}
We have computed a representative grid of evolutionary white dwarf models
of low mass for solar composition whose structure is consistent with their
earlier history of being remnants of the first giant branch. Though such models
have already been presented in the literature for isolated cases, our 
calculations are the first that cover a whole range of masses. 
These models allowed then to draw several important conclusions: 
  \begin{itemize}             
  \item

   We found thermal instabilities 
   due to hydrogen burning via the CNO cycle in a geometrically thin shell 
   (hydrogen shell flashes) at the beginning of
   the cooling branch of our evolutionary tracks.
   It turned out, however, that these shell flashes do occur only within a 
   limited mass range, $M=0.21\dots0.30~{\rm M}_{\odot}$, independently
   from the detailed input physics used.   
   \item

   The cooling speed is reduced by ongoing hydrogen burning at 
   the bottom of the envelope via the pp chains.
   This effect is the more pronounced the less massive the models are and may
   slow down the cooling down to 10\,000 K by up to a factor of about 40. 
   This is important for  age determinations of binary systems containing 
   helium-white dwarfs by using the cooling age of the white dwarf components.
   The employment of inconsistent white dwarf models may give grossly wrong 
   results in the sense that ages are underestimated.    
  
   We re-determined the age of the white dwarf component in 
   the \object{PSR J1012+5307} system with our evolutionary models and found
   an age of $6 \pm 1 \mathrm{Gyr}$, 15 times larger than that estimated from
   contraction models, but in excellent agreement with the pulsar's spin-down
   age of about 7~Gyr. We confirm the results of Alberts et al. (\cite{ASH}) 
   who also found 7~Gyr based on their white dwarf models.
  \item

  Our evolutionary models predict an inverse correlation of envelope mass
  with total mass, which is a continuation of the correlation already
  apparent for the more massive white dwarfs with carbon-oxygen cores 
  (Bl\"ocker et al. \cite{BHDB}). At the low mass end, the evolutionary 
  envelope masses are at least ten times larger than the often ad hoc assumed
  values for contraction models.
  Translated into a mass-radius relation,  
  these larger envelope masses lead to correspondingly
  higher mass estimates for any given $g,T_{\rm eff}$, and especially for the
  white dwarf companion of \object{PSR J1012+5307} a mass of 
  $M=0.19~{\rm M}_{\odot}$ follows, instead of $M=0.16 \,{\rm M}_{\odot}$ 
  by using existing contraction models.
  \end{itemize}
\begin{acknowledgements}
T.B. and F.H. acknowledge funding by the Deutsche
Forschungsgemeinschaft (grants Ko 738/12 and Scho 394/13).
We thank Detlev Koester for many helpful discussions on white dwarfs and
Marten van Kerkwijk for constructive remarks on the manuscript.
\end{acknowledgements}
%
%
%\bibliography{astro}

\begin{thebibliography}{astro}
%
\bibitem[1996]{ASH}
           Alberts F., Savonije G.J., van der Heuvel E.P.J., 
           1996, Nat. 380, 676
\bibitem[1994]{AF}
           Alexander D.R., Ferguson J.W., 1994, ApJ 437, 879
\bibitem[1997]{AB}
           Althaus L.G., Benvenuto O.G., 1997, ApJ 477, 313
\bibitem[1995]{BL}
           Bailes, M., Lorimer, D., 1995, in: Millisecond pulsars:
           a decade of surprises, A. S., Fruchter, M. Tavani, D. C. Backer (eds.),
           ASP Conf. Ser. 72
\bibitem[1998]{BA}
           Benvenuto, O. G., Althaus, L. G., 1998, MNRAS 293, 177
\bibitem[1995a]{Ba}
           Bl\"ocker T. 1995a, A\&A 297, 727
\bibitem[1995b]{Bb}
           Bl\"ocker T. 1995b, A\&A 299, 755
\bibitem[1997]{BHDB}
           Bl\"ocker T.,  Herwig F., Driebe T., Bramkamp H., 
           Sch\"onberner D., 1997, in: White Dwarfs, 
           Isern J., Hernanz M., Garcia-Berro E. (eds.), Kluwer, 
           Dordrecht, p.~57
\bibitem[1990]{BGRO}
           Bragaglia A., Greggio L., Renzini A., D'Odorico S.,
           1990, ApJ 365, L13
\bibitem[1996]{BKW}
           Burderi L., King A.R., Wynn G.A., 1996, MNRAS
           283, L63
\bibitem[1998]{CGK}
           Callanan, P. J., Garnavich, P. M., Koester, D., 1998, MNRAS 298, 207
\bibitem[1986]{CI}
           Cameron A. G. W., Iben I.\,Jr., 1986, ApJ 305, 228
\bibitem[1996]{C}
           Camilo F., 1996, in: {\it Pulsar Timing, General Relativity and 
           the Internal Structure of Neutron Stars}, 
           Proc.\ Coll.\ held at the Roy. Netherlands Academy of
           Arts \& Science, Eds.: Z. Arzoumanian, E. P. J. van den Heuvel \& 
           J. van Paradijs (Amsterdam: North Holland), in press.
\bibitem[1994]{CTK}
           Camilo F., Thorsett S. E., Kulkarni S. R., 1994, ApJ 421, L15
\bibitem[1996]{CEA}
           Camilo F., Nice D. J., Shrauner J. A., Taylor J. H., 1996,
           ApJ 469, 819 
\bibitem[1993]{CC}
           Castellani M., Castellani V., 1993, ApJ 407, 649 
\bibitem[1994]{CLR}
           Castellani V., Luridiana V., Romaniello M., 1994, ApJ 428, 633
\bibitem[1971]{ChSt}
           Chin C.-W., Stothers R., 1971, ApJ 163, 555
\bibitem[1965a]{CSa}
           Cox A.N., Stewart J.N., 1965a, ApJS 11, 1  
\bibitem[1965b]{CSb}
           Cox A.N., Stewart J.N., 1965b, ApJS 11, 22
\bibitem[1970]{CS70}
           Cox A.N., Stewart J.N., 1970, ApJS 19, 243
\bibitem[1994]{El}
           El Eid, M., 1994, A\&A 285, 915 
\bibitem[1996]{ES}     
           Ergma E., Sarna M.J., 1996, MNRAS 280, 1000
\bibitem[1970]{GRW}
           Gianonne P., Refsdal S., Weigert A., 1970, A\&A 4, 428
\bibitem[1961]{HS}
           Hamada T., Salpeter E.E., 1961, ApJ 134, 683           
\bibitem[1998a]{HPa}
           Hansen B.M.S., Phinney E.S., 1998a, MNRAS, 294, 557
\bibitem[1998b]{HPb}
           Hansen B.M.S., Phinney E.S., 1998b, MNRAS, 294, 569
\bibitem[1983]{IR83}
           Iben I.\,Jr., Renzini A., 1983, ARAA 21, 271
\bibitem[1984a]{IT84a}
           Iben I.\,Jr., Tutukov A.V., 1984a, ApJS 54, 335
\bibitem[1984b]{IT84b}
           Iben I.\,Jr., Tutukov A.V., 1984b, ApJ 282, 615
\bibitem[1986]{IT86}
           Iben I.\,Jr., Tutukov A.V., 1986, ApJ 311, 742
\bibitem[1995]{ITY95}
           Iben I.\,Jr., Tutukov A.V., Yungelson L.R.,
           1995, ApJS 100, 233
\bibitem[1996]{IR}
           Iglesias C.A., Rogers F.J., 1996, ApJ 464, 943
\bibitem[1992]{IRW}
           Iglesias C.A., Rogers F.J., Wilson B., 1992, ApJ 397, 717
\bibitem[1996]{KBK}
           van Kerkwijk M.H., Bergeron P., Kulkarni S.R., 1996, ApJ 467,
            L89           
\bibitem[1967]{KW}
           Kippenhahn R., Weigert A., 1967, Z.\ f.\ Astr.~65, 251
\bibitem[1967]{KKW}
           Kippenhahn R., Kohl K., Weigert, A., 1967,
           Z.\ f.\ Astr.~66, 58
\bibitem[1968]{KTW}
           Kippenhahn R., Thomas H.-C., Weigert A., 1968,
           Z.\ f.\ Astr.~69, 256
\bibitem[1986]{KS}
           Koester D., Sch\"onberner D., 1986, A\&A 154, 125
\bibitem[1995]{LFLN}
           Lorimer D.R., Festin L., Lyne A.G., Nicastro L., 1995, Nature 
           376, 393
\bibitem[1995]{Ly} Lyne, A.G. 1995, in: Compact stars in binaries,
IAU Symp. 165, E. P. J. van der Heuvel (ed.)
\bibitem[1996]{Ly1} Lyne, A.G. 1996, in: Proc. of the 7th M.\ Grossmann
Meeting, in press
\bibitem[1989]{MAME}
           Maeder, A., Meynet, G., 1989, A\&A 210, 155
\bibitem[1995]{MA}
           Marsh T.R., 1995, MNRAS 275, L1
\bibitem[1995]{MDD}
           Marsh T.R., Dillon V.S., Duck S.R., 1995, MNRAS 275, 828
\bibitem[1997]{MMB}
           Moran C., Marsh T.R., Bragaglia A., 1997, MNRAS 288, 538
\bibitem[1995]{NLLHBS}
           Nicastro L., Lyne A.G., Lorimer D.R., Harrison P.A., 
           Bailes M., Skidmore B.D., 1995, MNRAS, 273, L68
\bibitem[1994]{PK}
           Phinney E.S., Kulkarni S.R., 1994, ARA\&A 32, 591
\bibitem[1995]{RPJSH}
           Rappaport S., Podsiadlowski P., Joss P.C., Di Stefano R., 
           Han Z., 1995, MNRAS 273, 731
\bibitem[1996]{REA}
           Ray P. S., Thorsett S. E., Jenet F. A., et al. 1996,
           ApJ 470, 1103
\bibitem[1969]{RW}
           Refsdal S., Weigert A., 1969, A\&A 1, 167
\bibitem[1975]{R}
           Reimers D., 1975, Mem.\ Soc.\ Sci.\ Liege 8, 369
\bibitem[1988]{SLO}
           Saffer R.A., Liebert J., Olszewski E., 1988, ApJ 334, 947
\bibitem[1998]{SLY}
           Saffer R.A., Livio, M., Yungelson, L. R., 1998, ApJ, accepted
\bibitem[1996]{SMC}
           Sarna M.J., Marks P.B., Connon Smith R., 1996, MNRAS 279, 88
\bibitem[1979]{Sch79}
           Sch\"onberner D., 1979, A\&A 79, 108
\bibitem[1983]{Sch83}
           Sch\"onberner D., 1983, ApJ 271, 708
\bibitem[1994]{SL}Shore, S. N., Livio, M., van der Heuvel, E. P. J., 1994:
in {\it Interacting Binaries}, Saas-Fee Course 22, 
Swiss Soc. Astrophys. Astron., H.\ Nussbaumer \& A.\ Orr (eds.), Springer (Berlin)
\bibitem[1982]{SDW}
           Slattery W.L., Doolen G.D., DeWitt H.E., 1982, 
           Phys.\ Rev.\ A26, 2255
\bibitem[1996]{T}
           Tauris T.M., 1996, A\&A 315, 453
\bibitem[1996]{TB}
           Tauris T.M., Bailes M., 1996, A\&A 315, 432
\bibitem[1993]{VW93}
           Vassiliadis E., Wood P.R., 1993, ApJ 413, 641
\bibitem[1994]{VW94}
           Vassiliadis E., Wood P.R., 1994, ApJS 92, 125
\bibitem[1993]{V}
           Verbunt F., 1993, ARA\&A 31, 93
\bibitem[1975]{W75}
           Webbink R.F., 1975, MNRAS 171, 555
\bibitem[1984]{W84}
           Webbink R.F., 1984, ApJ 277, 355
\bibitem[1983]{WRS}
           Webbink R.F., Rappaport S., Savonije G.J., 1983, ApJ 270, 678
\bibitem[1994]{Wma}
           Wood M.A., 1994, in {\it White Dwarfs}, 
           D.\ Koester \& K.\ Werner (eds.),
           Springer, Berlin, p.~41
\end{thebibliography}
%
%

\end{document}